\documentstyle[12pt,seceqn,epsfig]{article}

\newcommand{\andvol}[3]{{\bf #1}~(#3)~#2}
\newcommand{\PRL}[3]{Phys.~Rev.~Lett.~\andvol{#1}{#2}{#3}}
\newcommand{\PRD}[3]{Phys.~Rev.~\andvol{D#1}{#2}{#3}}

\newcommand{\NPB}[3]{Nucl.~Phys.~\andvol{B#1}{#2}{#3}}
\newcommand{\PLB}[3]{Phys.~Lett.~\andvol{B#1}{#2}{#3}}

\newcommand{\ZPC}[3]{Z.~Phys.~\andvol{#1}{#2}{#3}}

\newcommand{\NCA}[3]{Il~Nuovo~Cim.~\andvol{#1A}{#2}{#3}}

\newcommand{\FP}[3]{Fortschr.~Phys.~\andvol{#1}{#2}{#3}}
\newcommand{\ND}{N_{\rm D}}
\newcommand{\NTC}{N_{\rm TC}}
\newcommand{\fTC}{f_{\rm TC}}
\newcommand{\fv}{f_{\rm V}}
\newcommand{\fa}{f_{\rm A}}
\newcommand{\mv}{m_{\rm V}}
\newcommand{\ma}{m_{\rm A}}
\newcommand{\Gamv}{\Gamma_{\rm V}}
\newcommand{\Gama}{\Gamma_{\rm A}}
\newcommand{\A}[1]{\Pi_{\rm #1}}
\newcommand{\B}[1]{\Pi'_{\rm #1}}

\newcommand{\F}[1]{{\cal F}_{\rm #1}}
\newcommand{\ImF}[1]{{\rm Im}\F{#1}}
\newcommand{\Jmu}[1]{J_{\rm #1}^\mu}
\newcommand{\Jnu}[1]{J_{\rm #1}^\nu}
\renewcommand{\S}[1]{S_{\rm #1}}

\newcommand{\Del}[1]{\Delta_{\rm #1}}
\newcommand{\polys}[1]{p_N(#1)}
\newcommand{\norm}[1]{\parallel #1 \parallel}
\newcommand{\beq}{\begin{equation}}
\newcommand{\eeq}{\end{equation}}
\newcommand{\beqx}{\begin{displaymath}}
\newcommand{\eeqx}{\end{displaymath}}
\newcommand{\beqa}{\begin{eqnarray}}
\newcommand{\eeqa}{\end{eqnarray}}
\newcommand{\beqax}{\begin{eqnarray*}}
\newcommand{\eeqax}{\end{eqnarray*}}
\begin{document}

\begin{titlepage}

%\begin{flushleft}
%{\bf Draft as of \today}
%\end{flushleft}

\begin{flushright}
UCTP--3/97 \\
VPI--IPPAP--97--3 \\
hep--ph/9702440
\end{flushright}

\bigskip
\bigskip
\bigskip

\begin{center}
{\Large\bf\boldmath ACD Estimation of the $S$-parameter Revisited}\\
\bigskip
\bigskip
{\normalsize
S.~R.~Ignjatovi\'c$^{1)}$,
T.~Takeuchi$^{2)}$, and L.~C.~R.~Wijewardhana$^{1)}$ \\
\medskip
$^{1)}${\it 
Department of Physics, University of Cincinnati, Cincinnati, OH 45221} \\
\smallskip
$^{2)}${\it 
Institute for Particle Physics and Astrophysics, \\
Physics Department, Virginia Tech, Blacksburg, VA 24061-0435
} 
}
\end{center}

\bigskip
\bigskip
\bigskip

\begin{abstract}
The analytic continuation by duality (ACD) technique has been used to
estimate the electroweak $S$ parameter in technicolor models.
In this letter, we investigate the reliability of this method by applying it
to some toy models with known spectra. 
We find that in most instances the technique cannot be trusted to 
give a reliable result.
\end{abstract}

\bigskip

\vfill

\begin{flushleft}
UCTP--3/97 \\
VPI--IPPAP--97--3 \\
hep--ph/9702440 \\
February 1997
\end{flushleft}

\end{titlepage}

%%%%%%%%%%%%%%%%%%%%%%%%%%%%%%%%%%%%%%%%%%%%%%%%%%%%%%%%%%%%%%%%%%%%%%%%%%%%%%%
%%%%%%%%%%%%%%%%%%%%%%%%%%%%%%%%%%%%%%%%%%%%%%%%%%%%%%%%%%%%%%%%%%%%%%%%%%%%%%%
\section{Introduction}

There is no need to emphasize the importance of
finding reliable non--perturbative calculational techniques 
that can be applied to strongly interacting theories such as technicolor.
Without such methods, strongly interacting theories can never be
tested since no quantitative predictions that can be compared 
directly with experiment can be made.

The {\it analytic continuation by duality} (ACD) technique, which
was first developed in Ref.~\cite{nasrallah},
has been proposed in Ref.~\cite{SH} 
as a potentially reliable way to 
compute the oblique correction parameter 
$S$ for technicolor theories.
The advantage of the ACD technique was that it could be
applied to both QCD--like and walking technicolor \cite{walking}
theories whereas the dispersion relation technique used by 
Peskin and one of us
in Ref.~\cite{peskin} could only be applied to the former.
Furthermore, the ACD estimate of $S$ for walking technicolor
implied that walking dynamics could render $S$ negative,
making it compatible with the current experimental limit.
This was in contrast to the result of Harada and Yoshida \cite{HY}
who used the Bethe--Salpeter equation approach to conclude that
$S$ was positive even for walking theories.

However, the robustness of the ACD technique has been 
questioned in the literature \cite{caprini}.
In this letter, we further investigate the reliability of the ACD technique.
In section 2, we review the definition of the $S$ parameter and 
explain the ACD technique.
In section 3, we apply the ACD technique to the perturbative spectral
function that represents the contribution of a degenerate doublet of heavy
fermions to see if the well known result $1/6\pi$ could be reproduced.
In section 4, we apply the ACD technique to a model function
which has an imaginary part more representative of the actual QCD
spectrum.   Section 5 concludes.

\section{The ACD technique}

The ACD technique consists
of using the analyticity of the vacuum polarization function to convert the
dispersion integral  for S, which is an integral along
the real $s$ axis, into an integral  around a large circle in the complex  $s$
plane. The value of the integral on the circle is then estimated by
 the Operator Product Expansion (OPE) \cite{shifman}. 

We denote by $\A{XY}$ the $g^{\mu\nu}$ part of the vacuum polarization
tensor between two currents $\Jmu{X}$ and $\Jnu{Y}$:
\beqx
ig^{\mu\nu}\A{XY}(s) + (q^\mu q^\nu {\rm term})
= \int d^4x\;e^{iq\cdot x}\langle \Jmu{X}(x)\Jnu{Y}(0)
                          \rangle,
\qquad s=q^2.
%\label{AXYDEF}
\eeqx
The $S$ parameter is defined in Ref.~\cite{peskin} as 
\beqx
S = -4\pi\left[ \B{VV}(0) - \B{AA}(0)
         \right]
%\label{SDEF}
\eeqx
where the subscripts $V$ and $A$ denote the neutral
isospin--1 vector and axial--vector currents, 
respectively, and the prime denotes a derivative
with respect to $s$.   

For latter convenience, we define
the functions $\F{V}(s)$ and $\F{A}(s)$ as
\beqax
\A{VV}(s) & \equiv & s\F{V}(s), \cr
\A{AA}(s) & \equiv & \A{AA}(0) + s\F{A}(s),
\eeqax
and the following shorthand notation for
the difference between $\F{V}(s)$ and $\F{A}(s)$:
\beqx
\F{}(s) \equiv \F{V}(s)  - \F{A}(s).
\eeqx
Then,
\beqx
S = -4\pi[\F{V}(0) - \F{A}(0)] = -4\pi\F{}(0).
\eeqx
Note that our notation is slightly different from either
Ref.~\cite{SH} or Ref.~\cite{peskin} so care is necessary when 
comparing formulae.

\begin{figure}[t]
\begin{center}
\epsfig{file=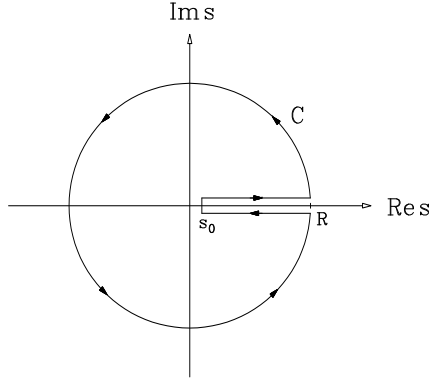,angle=90,height=5cm,width=5.6cm}
\caption{The contour $C$ which avoids the branch cut along the
           real $s$--axis.}
\label{CONTOUR}
\end{center}
\end{figure}

The function $\F{}(s)=\F{V}(s)-\F{A}(s)$ is analytic in the entire
complex $s$ plane except for a branch cut along the positive real $s$ 
axis starting from the lowest particle threshold contributing to
$\F{}(s)$.   Therefore, applying Cauchy's theorem to the contour 
$C$ shown in Fig.~\ref{CONTOUR}, we find
\beqa
S & = & -4\pi\F{}(0)   \cr
  & = & -4\pi
        \left[ \frac{1}{\pi}\int_{s_0}^R ds \frac{ \ImF{}(s) }{ s }
             + \frac{1}{2\pi i}\oint_{|s|=R} ds \frac{ \F{}(s) }{ s }
        \right].
\label{CAUCHY1}
\eeqa
If the radius of the contour $R$ is taken to infinity, the integral
around the circle at $|s|=R$ vanishes and we obtain the dispersion
relation
\beqx
S = -4\int_{s_0}^R ds \frac{ \ImF{}(s) }{ s },
\eeqx
which was used in Ref.~\cite{peskin} to calculate $S$.
However, the dispersion relation approach requires the knowledge of
$\ImF{}(s)$ along the real $s$ axis
which is only available for QCD--like theories.

The basic idea of the ACD technique, on the other hand, 
is to approximate the kernel $1/s$ by a polynomial
\beqx
\frac{1}{s} \approx \polys{s} = \sum_{n=0}^{N} a_n(N) s^n,
\qquad s\in[s_0,R]
\label{FIT}
\eeqx
and use it to make the integral along the real $s$ axis vanish instead.
Applying Cauchy's theorem to the product $p_N(s)\F{}(s)$ over the 
same contour $C$ yields 
\beq
0 = -4\pi\left[ \frac{1}{\pi}\int_{s_0}^R \!ds\,\polys{s}\ImF{}(s)
              + \frac{1}{2\pi i}\oint_{|s|=R}\!ds\,\polys{s}\F{}(s)
         \right].
\label{ZERO}
\eeq
Subtracting Eq.~(\ref{ZERO}) from Eq.~(\ref{CAUCHY1}), we obtain
\beqx
S = S_N + \Del{fit},
\eeqx
where
\beqax
S_N & \equiv & - \frac{2}{i}\oint_{|s|=R} \!ds
                 \left[ \frac{1}{s} - \polys{s} \right] \F{}(s), \cr
\Del{fit}
   & \equiv & -4\int_{s_0}^R \!ds 
                \left[ \frac{1}{s} - \polys{s} \right] \ImF{}(s).
\eeqax
For sufficiently large $N$, we can expect $\Del{fit}$ to
be negligibly small.  In fact, it converges to zero in the
limit $N\rightarrow \infty$ (though how quickly the convergence
occurs depends on the interval $[s_0,R]$).
We can therefore neglect it and approximate
$S$ with $S_N$ which is an integral around the circle
$|s|=R$ only.   We call $\Del{fit}$ the {\it fit error}.

In fitting $\polys{s}$ to $1/s$, 
the $s$--coefficients $a_n(N)$'s are determined so that some 
{\it norm} of the difference function
\beqx
d(s)=\frac{1}{s}-\polys{s}
\eeqx
on $[s_0,R]$ is minimized. It is usually an $L_p$ norm:
\beqx
\norm{d} \equiv \left[ \int_{s_0}^R |d(s)|^p w(s) ds
                \right]^{1/p},
\eeqx
where $w(s)$ is some weight function. 
Obviously, different choices of $p$ and
$w(s)$ will lead to different fitting coefficients $a_n(N)$,
thus different values for $S_N$ and $\Del{fit}$.
Since this ambiguity has not been explored in the literature on ACD, 
we have done a detailed analysis that will be presented elsewhere \cite{new}. 
The fitting routine dependence proved to be mild.

If the radius of the contour $R$ is taken to be sufficiently large,
the function $\F{}(s)$ can be approximated on $|s|=R$ 
by a large momentum expansion: 
\beq
\F{}(s) \approx \sum\limits_{m=1}^M \frac{h_m(s)}{s^m}.
\label{ope}
\eeq
This expression is obtained by analytically continuing the 
operator product expansion (OPE)
of $\F{}(s)$ \cite{shifman}, which is available for both QCD--like and
walking technicolor theories, from the deep Euclidean region $s \ll 0$.
Therefore, we can write
\beqx
S_N = S_{N,M} + \Del{tr},
\eeqx
where
\beqax
S_{N,M}  & \equiv & -\frac{2}{i}\oint_{|s|=R} \!ds
                     \left[ \frac{1}{s} - \polys{s}
                     \right]
                     \sum_{m=1}^M \frac{h_m(s)}{s^m}, \cr
\Del{tr} & \equiv & -\frac{2}{i}\oint_{|s|=R} \!ds
                     \left[ \frac{1}{s} - \polys{s}
                     \right] 
                     \left[ \F{}(s)
                          - \sum_{m=1}^M \frac{h_m(s)}{s^m}
                     \right],
\label{trunc}
\eeqax
and approximate $S_N$ with $S_{N,M}$.  The neglected term
$\Del{tr}$ is called the {\it truncation error}.
It is actually not clear whether the truncation error will go to
zero in the limit $M\rightarrow\infty$ since Eq.~(\ref{ope})
may be an asymptotic series and not a convergent one.
If it is asymptotic, taking $M$ to be larger than an optimum
value may worsen the approximation.  However, $M$ must be taken to be
larger than $N$ to suppress $\Del{tr}$ with inverse powers of $R$.
If the series is convergent, the convergence may not
be particularly fast requiring a very large $M$ and/or $R$ to suppress
$\Del{tr}$.   Some of these points will be discussed in the following,
but a more complete treatment will be given in Ref.~\cite{new}.

If the $s$--dependence of the 
expansion coefficients in Eq.~(\ref{ope}) is 
``negligibly weak'' then 
the approximation can be taken one step further and the 
$s$--dependence of the $h_m(s)$'s dropped,
{\it i.e.}
\beqx
h_m(s) \approx h_m(-R) \equiv \hat{h}_m.
\eeqx
(One must keep in mind that this is a very dangerous approximation 
to make since the analytic structure of the integrand will be completely 
altered no matter how ``weak'' the $s$--dependence may be: the imaginary part
of $\F{}(s)$ will be reduced to derivatives of 
$\delta$--functions at the origin.) Therefore,
\beqx
S_{N,M} = \S{ACD} + \Del{AC}
\eeqx
where
\beqax
\S{ACD} & \equiv & -\frac{2}{i}\oint_{|s|=R}\!ds
               \left[ \frac{1}{s} - \polys{s}
               \right] \sum_{m=1}^M \frac{\hat{h}_m}{s^m}, \cr
\Del{AC} & \equiv & -\frac{2}{i}\oint_{|s|=R}\!ds
               \left[ \frac{1}{s} - \polys{s}
               \right] \sum_{m=1}^M \frac{h_m(s)-\hat{h}_m}{s^M},
\eeqax
and we can approximate $S_{N,M}$ with $S_{ACD}$.
Unfortunately, this is a tautological statement: 
The only way one can conclude that the $s$--dependence was 
``negligibly weak'' is to calculate $\Del{AC}$ and find that
it is indeed suppressed.
However, it is possible to give the following handwaving argument
that $\Del{AC}$ would be generically small: 
The difference $h_m(s)-\hat{h}_m$ in the integrand of $\Del{AC}$ 
can be expected to be smallest in the deep Euclidean region and most 
pronounced in the vicinity of the positive
real $s$ axis.   But since that is where the difference 
$1/s-\polys{s}$ is approximately zero, the contribution of this
large difference to $\Del{AC}$ with be highly suppressed.
Thus:
\beqx
S \approx \S{ACD}.
\eeqx
In this approximation, the integral for $\S{ACD}$ will only pick
up the residues of the single poles inside the integration contour
and we find,
\beqx
\S{ACD} = 4\pi\sum\limits_{n=0}^{\min\{N,M-1\}}a_n(N) \hat{h}_{n+1}.
\label{Sacd}
\eeqx
We will call $\Del{AC}$ the {\it analytical continuation error}. 

To summarize, the ACD technique uses the relation
\beqx
S = \S{ACD} + \Del{AC} + \Del{tr} + \Del{fit},
\eeqx
and assumes that all three types of error can be neglected and
approximates $S$ with $\S{ACD}$.
Therefore, the question of whether the ACD technique
is reliable or not can be translated to the question of
whether all these errors are under control or not.

In actual applications of the ACD technique, 
there is an additional source of error due to the fact that the
expansion coefficients $\hat{h}_m$ are only known approximately.
This is because the $\hat{h}_m$'s are products of Wilson coefficients
and condensates of operators.
The Wilson coefficients can only be calculated perturbatively,
and only a very limited number of them are known.  The condensates
are even less known and only a handful of them have been determined
phenomenologically.
As a result, the numerical uncertainty on $\hat{h}_m$ increases
with $m$ so increasing $N$ and $M$ does not necessarily improve
the final estimate of $\S{ACD}$ since it may be swamped by the
numerical uncertainty of the higher order terms one must include.
A common strategy in the literature is to try to choose optimum values
of $N$ and $M$ such that the fit and truncation errors are suppressed,
but the numerical uncertainty is not yet too large.
The procedure essentially consists of the minimization of a `total' error.
However, it is very difficult to define the total error since it involves a
mathematically well defined error and, on the other hand, an error which
reflects our ignorance.   The problem itself is ill--posed and
difficult to treat with any level of rigor.

We will therefore ignore this last type of error 
in this letter and confine ourselves to the
question: {\it if\/} the expansion coefficients of Eq.~(\ref{ope}) were
known {\it exactly}, would the ACD technique give the correct
result?

\section{The Perturbative Spectral Function}

\begin{table}[t]
\caption{$\S{ACD}$ and the fit, truncation, and analytical continuation
errors for the perturbative model function. 
The cutoffs are $[s_0,R] = [4m^2,25m^2]$, and the fit routine was the
least square fit.  
The exact value of $S$ is $1/6\pi = 0.0531$.}
\begin{center}
\begin{tabular}{cccccc}
\hline
\hline
$N$ & $M$ & $\S{ACD}$ & $S_{N,M}=\S{ACD}+\Del{AC}$ & $\Del{fit}$ & $\Del{tr}$
 \\
\hline\hline
$3$ & $2$ &  $0.2930$ &  $0.0580$ & $-0.0002$ & $-0.0048$  \\
    & $3$ &  $0.2883$ &  $0.0530$ &           & $\phantom{-}0.0002$ \\
    & $4$ &  $0.2884$ &  $0.0532$ &           & $-0.0000$  \\
\hline
$4$ & $2$ &  $0.4330$ &  $0.0632$ & $-0.0001$ & $-0.0101$  \\
    & $3$ &  $0.4203$ &  $0.0521$ &           & $-0.0010$  \\
    & $4$ &  $0.4211$ &  $0.0532$ &           & $-0.0000$  \\
    & $5$ &  $0.4211$ &  $0.0531$ &           & $\phantom{-}0.0000$ \\
\hline
$5$ & $3$ &  $0.5731$ &  $0.0506$ & $-0.0000$ & $\phantom{-}0.0025$ \\
    & $4$ &  $0.5759$ &  $0.0533$ &           & $-0.0002$  \\
    & $5$ &  $0.5757$ &  $0.0531$ &           & $\phantom{-}0.0000$ \\
    & $6$ &  $0.5757$ &  $0.0531$ &           & $-0.0000$  \\
\hline\hline
\end{tabular}
\end{center}
\label{tab1}
\end{table}

Now we investigate the reliability of the ACD technique by evaluating
$\S{ACD}$ and the three errors 
$\Del{fit}$, $\Del{tr}$, and $\Del{AC}$
in some toy models. 
First we take the one--loop perturbative contribution of a
doublet of heavy fermions to $S$. 
The contribution to the vacuum polarization coming from 
this fermion doublet can readily be computed \cite{peskin}, 
and therefore the $S$ parameter can easily be evaluated; 
its value is $1/6\pi$.

Next, we use the ACD technique to do the same computation.
The perturbative spectral function is
\beqx
\F{pert}(s) =
\frac{1}{4\pi^2}\frac{m^2}{s}
\int_0^1 \!dx\log\left[ 1 - x(1-x)\frac{s}{m^2}
                 \right].
%\label{FPERT}
\eeqx
It is analytic in the entire complex $s$ plane except for a branch cut
along the positive real $s$ axis which starts from $s=4m^2$.
The imaginary part of this function along the cut is given by
\beqx
\ImF{pert}(s) = -\frac{1}{4\pi}
                 \frac{m^2}{s}\beta
                 \theta(s-4m^2),
\qquad
\beta = \sqrt{1-\frac{4m^2}{s}}.
%\label{IMFPERT}
\eeqx
The first few terms of the large $s$-expansion of $\F{pert}(s)$ are given by
\beqax
\lefteqn{-4\pi^2\F{pert}(s)} \cr
& = & x
      \left\{ -\ln\left(-\frac{1}{x}\right)+2
      \right\}
    + x^2 
      \left\{ 2\,\ln\left(-\frac{1}{x}\right)+2
      \right\}
    + x^3 
      \left\{ 2\,\ln\left(-\frac{1}{x}\right)-1
      \right\}
+ \dots,
\eeqax
where $x\equiv 4m^2/s$.
Using these expressions, we calculated $\S{ACD}$,
$\Del{AC}$, $\Del{tr}$, and $\Del{fit}$.
The result of our calculations for several values of $N$ and $M$
are shown in Table~\ref{tab1}.
The fit interval was $[s_0,R]=[4m^2,25m^2]$,
and the fit routine was the least square fit.

As is evident from Table~\ref{tab1},
the fit and truncation errors are under excellent control and
$S_{N,M}$ reproduces the exact value of $S$ accurately
already at $N=M=3$.
However, the analytic continuation error is not.
For the $N=5$ case, for instance,  $\S{ACD}$ is larger 
than the exact value by more than an order of magnitude.
In fact, we find that $\S{ACD}$ and $\Del{AC}$ diverge 
as $N\rightarrow\infty$.   

We conclude that neglecting the $s$--dependence
of the $h_m(s)$'s fails miserably as an approximation.
The reason for this can be traced to the fact that even 
though the difference $1/s - \polys{s}$ converges to zero within
its radius of convergence, outside it diverges.
Therefore, the handwaving argument we gave in the previous section was wrong:
the error induced by the neglect of the $s$--dependence of the
$h_m(s)$'s may be suppressed near the real $s$ axis, but it is actually
{\it enhanced} away from it.

\section{The Breit--Wigner Model}

\begin{figure}[ht]
\begin{center}
%\vspace{7cm}
\epsfig{file=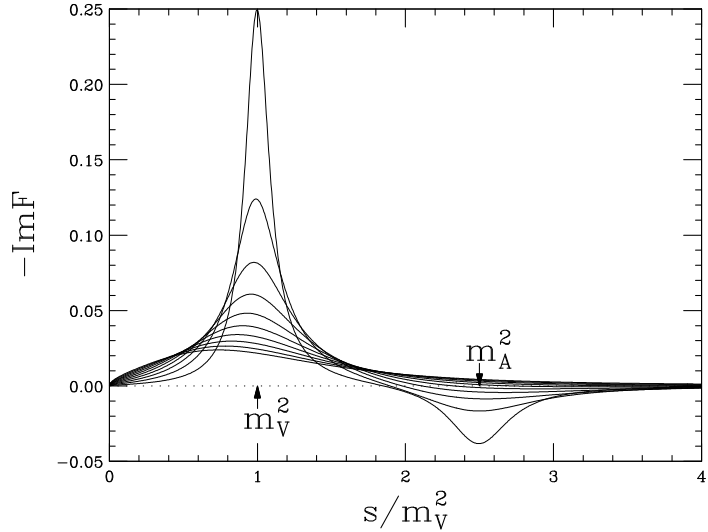,angle=90,height=7cm}
\caption{The Spectral Functions for the Breit--Wigner Model.
The ratio $\xi \equiv \Gamv/\mv = \Gama/\ma$ is varied from 
$\xi=1.0$ to $\xi=0.1$ at $0.1$ intervals with smaller
values of $\xi$ corresponding to more prominent peaks.
The arrows indicate the positions of the techni--$\rho$ 
and techni--$a_1$ peaks.}
\label{RVMD}
\end{center}
\end{figure}

In light of our result for the perturbative spectral function,
the natural question to ask next is whether including the 
$s$--dependence of the expansion coefficients, {\it i.e.} not
neglecting $\Del{AC}$, would give an accurate estimate of
$S$ for all cases.

To answer this question, we consider the following
model:
\beqax
\F{V}(s) & = & \fv^2 \frac{1}{s-\mv^2+i\sqrt{s}\Gamv}, \cr
\F{A}(s) & = & \fa^2 \frac{1}{s-\ma^2+i\sqrt{s}\Gama}.
\eeqax
These functions are analytic in the complex $s$ plane except for
a branch cut along the positive real $s$ axis starting from $s=0$.
The imaginary part along the cut is of the Breit--Wigner type:
\beqax
\ImF{V}(s) 
& = & -\fv^2 \frac{ \sqrt{s}\Gamv }
                  { (s-\mv^2)^2 + s\Gamv^2 }\theta(s), \cr
\ImF{A}(s)
& = & -\fa^2 \frac{ \sqrt{s}\Gama }
                  { (s-\ma^2)^2 + s\Gama^2 }\theta(s). 
\eeqax

We impose the Weinberg sum rules \cite{weinberg} on these functions
which leads to 
the constraints\footnote{Actually, the integral for the second Weinberg 
sum rule does not converge for this model.   We therefore impose the
equivalent constraint that the $1/Q^4$ term vanishes in the OPE
for $\F{}(-Q^2)$.}
\beqax
\fv^2 - \fa^2 & = & \fTC^2, \cr
\fv^2\mv^2 - \fa^2\ma^2 & = & 0,  \cr
\fv^2\Gamv^2 - \fa^2\Gama^2 & = & 0,
\eeqax
where $\fTC$ is the technipion decay constant.

The value of $S$ for this model is given by
\beqx
S = -4\pi\left[ \F{V}(0) - \F{A}(0)
         \right]
  = 4\pi\left[ \frac{\fv^2}{\mv^2} - \frac{\fa^2}{\ma^2}
        \right]
  = 4\pi(1+r)\frac{\fTC^2}{\mv^2}
\eeqx
where $r\equiv\mv^2/\ma^2$.  We fix $r$ and the ratio $\fTC^2/\mv^2$
using large--$N$ rescaling from QCD:
\beqax
\frac{\fTC^2}{\mv^2} & \approx &
\frac{f_\pi^2}{m_\rho^2}\frac{\ND\NTC}{3}
\;\approx\; 0.005\ND\NTC, \cr
r \;=\; \frac{\mv^2}{\ma^2} & \approx &
\frac{m_\rho^2}{m_{a_1}^2} \;\approx\; 0.4
\eeqax
so that
\beqx
S \approx 0.09\ND\NTC.
\eeqx

Note that the value of
$S$ does not depend on the value of the width to mass ratio
\beqx
\xi \equiv \frac{\Gamv}{\mv} = \frac{\Gama}{\ma}.
\eeqx
However, the spectral function and the $s$ dependence of the
large momentum expansion coefficients do.
In Fig.~\ref{RVMD}, we show the $\xi$--dependence of $\ImF{}(s)$.
In the limit $\xi \rightarrow +0$, these functions reduce to the
well known $\delta$--function (vector-meson dominance) model,
and the large momentum expansion coefficients have no $s$--dependence.
In the other limit where $\xi \approx 1$, 
we obtain a smooth function reminiscent of the
perturbative case, and the $s$--dependence of the
expansion coefficients is large.
In the following, we fix $\xi$ to the large-$N$ value
\beqx
\xi = \frac{\Gamv}{\mv} \approx \frac{\Gamma_\rho}{m_\rho} \approx 0.2,
\eeqx
which is in between the two extremes.

\begin{table}[ht]
\begin{center}
\caption{The $S$-parameter for the Breit-Wigner model function. 
The cutoffs are $s_0 = 0.2m_V^2$ and $R = 5m_V^2$. 
The dimensionless parameters are $r= 0.4$, $\xi = 0.2$. 
The exact value of $S$ is $0.257$.
(Since the $\S{IR}=0.016$ correction is not included here, the results
should be compared with $0.241$).
}
\smallskip
\begin{tabular}{ccrcrr}
\hline
\hline
$N$ &   $M$  & $\S{ACD}$  &   $S_{N,M} = \S{ACD} + \Del{AC}$
    & $\Del{fit}$   &   $\Del{tr}$  \\
\hline\hline
$3$ &   $2$  &   $0.657$  & $\phantom{-}0.557$ &   $-0.057$ &  $-0.244$ \\
    &   $3$  &   $0.365$  & $\phantom{-}0.102$ &            &   $0.211$ \\
    &   $4$  &   $0.375$  & $\phantom{-}0.343$ &            &  $-0.030$ \\
    &   $5$  &   $0.375$  & $\phantom{-}0.305$ &            &   $0.009$ \\
\hline
$4$ &   $2$  &   $0.838$  & $\phantom{-}0.836$ &   $-0.006$ &  $-0.573$ \\
    &   $3$  &  $-0.063$  &           $-0.445$ &            &   $0.708$ \\
    &   $4$  &   $0.009$  & $\phantom{-}0.624$ &            &  $-0.361$ \\
    &   $5$  &   $0.118$  & $\phantom{-}0.183$ &            &   $0.080$ \\
    &   $6$  &   $0.118$  & $\phantom{-}0.244$ &            &   $0.019$ \\
\hline
$5$ &   $3$  &  $-1.083$  &           $-1.578$ &    $0.014$ &   $1.821$ \\ 
    &   $4$  &  $-0.811$  & $\phantom{-}1.604$ &            &  $-1.361$ \\
    &   $5$  &   $0.093$  & $\phantom{-}0.402$ &            &   $0.645$ \\
    &   $6$  &  $-0.230$  & $\phantom{-}0.327$ &            &  $-0.085$ \\
    &   $7$  &  $-0.230$  & $\phantom{-}0.238$ &            &   $0.005$ \\
\hline
$8$ &   $7$  &  $11.687$  &           $-6.149$ &    $0.000$ &   $6.406$ \\
    &   $8$  &  $-3.111$  & $\phantom{-}2.397$ &            &  $-2.141$ \\
    &  $10$  &  $-1.195$  & $\phantom{-}0.028$ &            &   $0.234$ \\
\hline\hline
\end{tabular}
\label{tab2}
\end{center}
\end{table}

The large momentum expansion of our model function
is given by
\beqx
\F{}(s)
     = \frac{\fTC^2}{s}\sum_{n=0}^\infty
      \frac{(-1)^n \mv^{2n}}{s^n}
      \left[ X_{2n}(r) U_{2n}(\xi/2)
           -i\frac{ \mv }{ \sqrt{s} }
             X_{2n+1}(r) U_{2n+1}(\xi/2)
      \right]
%\label{BWOPE}
\eeqx
where $X_{n}(r)$ is defined as
\beqx
X_n(r) \equiv \frac{1-r^{1-\frac{n}{2}}}{1-r},
\eeqx
while $U_n(\cos\theta) = \sin(n+1)\theta/\sin\theta$ are 
the Chebyshev polynomials of the second kind.
Note that in the limit $\xi=0$, the $s$--dependence of the
coefficients vanishes because $U_{2n+1}(0)=0$.

We have a slight problem with the choice of fit interval
since the branch cut of this model begins at $s=0$
but the IR cutoff $s_0$ must be kept non--zero.
We will therefore choose $s_0$ to be well below the 
peak of the techni--$\rho$ 
resonance but non-zero. The approximate
form of the spectral function below $s_0$ is
\beqx
-\ImF{}(s)\;\approx\; \left( \frac{\fv^2}{\mv^3} - \frac{\fa^2}{\ma^3}
                      \right)\xi\sqrt{s}
          \;   =   \; \frac{\xi\fTC^2 X_{-3}(r)\sqrt{s}}{\mv^3}.
\eeqx
This can be used to calculate analytically the contribution from the 
interval $[0,s_0]$:
\beqx
\S{IR}(s_0)\;=\; -4\int_0^{s_0} \frac{ds}{s} \ImF{}(s)
\;\approx\;       \frac{8\xi\fTC^2 X_{-3}(r)\sqrt{s_0}}{\mv^3}.
\eeqx
The ACD technique is employed for estimating the contribution
coming from the interval $[s_0,R]$. 
We must modify the large momentum expansion of $\F{}(s)$ slightly to
subtract out the contribution of the interval $[0,s_0]$:
\beqx
\F{}(s) \Longrightarrow \F{}(s) - \delta\F{IR}(s)
\eeqx
where
\beqax
\delta\F{IR}(-Q^2) & = & \frac{1}{\pi}\int_0^{s_0}ds
                         \frac{\ImF{}(s)}{s+Q^2}   \cr
& \approx & \left[\frac{\xi\fTC^2 X_{-3}(r)\sqrt{s_0}}{\pi\mv^3}
            \right]
            \sum_{m=1}^\infty \frac{(-1)^m}{(m+\frac{1}{2})}
                              \frac{s_0^m}{Q^{2m}},
\eeqax
and our estimate of $S$ will be given by
\beqx
\S{} \;\approx\; \S{IR}(s_0)+\S{ACD}(s_0,R).
\eeqx
For our present analysis, we used the values 
$s_0/\mv^2 = 0.2$, and $R/\mv^2 = 5$. 

Using these numbers and expressions, we calculated $\S{ACD}$
and the three errors $\Del{AC}$, $\Del{tr}$, and $\Del{fit}$
for the case $\ND=1$, $\NTC=3$.
The results are shown on Table~\ref{tab2}. 
In this case,
not only did the $\S{ACD}$
approximation fail, but even the inclusion of the $s$-dependence did not
give accurate and stable results. 
The reason is that the truncation error 
is oscillatory and converges too slowly.
It will eventually converge since the large momentum expansion is
convergent for this model, but not before $M$ becomes impractically large;
in a realistic application one would have to truncate the OPE at $M = 3$ or 
$4$.

\section{Conclusions}

We have shown that the ACD technique employed by Sundrum and Hsu in 
estimating the $S$-parameter is not completely reliable. 
We find that the neglect of the $s$-dependence of OPE coefficients
is not a good approximation.
In an example that we have analyzed we find that the
truncation error is also not controllable, even if the spectral function is
exactly known. 
This brings into doubt the utility of the ACD method in
estimating the $S$ parameter in complicated theories like walking technicolor.

\section*{Acknowledgments}

We would like to thank R.~Sundrum for helpful discussions.
S.~R.~I. and L.~C.~R.~W. were supported in part by the U.S. Department of
Energy under the grant \#DE--FG02--84ER40153.


\begin{thebibliography}{10}

\newcommand{\brk}{\hfill\\}

\bibitem{nasrallah} 
{N.~F.~Nasrallah, N.~A.~Papadopoulos and K.~Schilcher, \brk
 \PLB{113}{61}{1982}; \andvol{B126}{379}{1983}; \ZPC{16}{323}{1983}. \brk
 N.~A.~Papadopoulos, J.~A.~Pe\~narocha, F.~Scheck and K.~Schilcher, \brk
 \PLB{149}{213}{1984}; \NPB{258}{1}{1985}.}

\bibitem{SH}
{R.~Sundrum and S.~D.~H.~Hsu, \NPB{391}{127}{1993}.}

\bibitem{walking}
{B.~Holdom, \PLB{150}{301}{1985}, \brk
 T.~Appelquist, D.~Karabali, and L.~C.~R.~Wijewardhana, \brk
 \PRL{57}{957}{1986};\brk
 T.~Appelquist and L.~C.~R.~Wijewardhana, \brk 
 \PRD{35}{774}{1987}, \PRD{36}{568}{1987};\brk
 K.~Yamawaki, M.~Bando and K.~Matumoto, \brk \PRL{56}{1335}{1986}.}

\bibitem{peskin} 
{M. E. Peskin and T. Takeuchi, \brk
 \PRL{65}{964}{1990}; \PRD{46}{381}{1992}; \brk
 T.~Takeuchi, in {\it the Proceedings of the International Workshop on
 Electroweak Symmetry Breaking}, Hiroshima, Japan, November 1991, edited by
 W.~A.~Bardeen, J.~Kodaira, and T.~Muta (World Scientific, Singapore, 1992).
 \brk
 The $S$ parameter has also been estimated by employing low energy effective
 Lagragian techniques by: \brk
 B.~Holdom and J.~Terning, \PLB{247}{88}{1990}, \brk
 M.~Golden and L.~Randall, \NPB{361}{3}{1991}.
}

\bibitem{HY}
{M.~Harada and Y.~Yoshida, \PRD{50}{6902}{1994}.}

\bibitem{caprini}
{I.~Caprini and C.~Verzegnassi, \NCA{80}{187}{1984}.}

\bibitem{shifman} 
{M.~A.~Shifman, A.~I.~Vainshtein, and V.~I.~Zakharov, \brk
 \NPB{147}{385}{1979}; \brk
 V.~A.~Novikov, M.~A.~Shifman, A.~I.~Vainshtein, and V.~I.~Zakharov, \brk
 \FP{32}{585}{1984}; \NPB{249}{445}{1985}.}

\bibitem{new}
{L.~C.~Goonetileke, S.~R.~Ignjatovi\' c, T.~Takeuchi and \brk 
 L.~C.~R.~Wijewardhana (in preparation).}

\bibitem{weinberg}
{S.~Weinberg, \PRL{18}{507}{1967}.}


\end{thebibliography}
\end{document}